\documentclass[a4paper,10pt,twoside]{cpc-hepnp}
\usepackage{CJK,upgreek,fancyhdr}
\usepackage{multicol}
\usepackage{graphicx}
\usepackage{booktabs}
\usepackage{amssymb,bm,mathrsfs,bbm,amscd}
\usepackage[tbtags]{amsmath}
\usepackage{lastpage}
\usepackage{subcaption}
\usepackage{textcomp}
\usepackage{url}
\usepackage{hyperref}
\usepackage{lipsum,float}
\usepackage[running]{lineno}

\newcommand{\tabincell}[2]{\begin{tabular}{@{}#1@{}}#2\end{tabular}} 
\begin{document}
\begin{CJK*}{GB}{gbsn}

\fancyhead[c]{\small }

\footnotetext[0]{}

\title{Study of linearity of LYSO crystal for HERD calorimeter\thanks{Supported by National Natural Science Foundation of China, Grant No.11327303, Grant No. 11473028 and Grant No. 11675196; International Science and Technology Cooperation Program of China, Grant No. 2015DFA10140; Interdisciplinary Innovation Team Project of Chinese Academy of Sciences (Research Team of The High Energy cosmic-Radiation Detection); Qianren start-up grant 292012312D1117210; Strategic Pioneer Program in Space Science, Chinese Academy of Sciences, Grant No. XDA04075600; Youth Innovation Promotion Association of CAS, Grant NO. 2014009 }}

\author{%
  Zheng Quan (权征)$^{a,b}$
\quad Zhi-gang Wang (王志刚)$^{c;1)}$\email{wangzhg@mail.ihep.ac.cn}
\quad Ming Xu (徐明)$^{a}$
\\\quad Yong-wei Dong (董永伟)$^{a}$
\quad Jun-jing Wang (王俊静)$^{a,b}$
\quad Guang-peng An (安广朋)$^{c}$
\\\quad Xin Liu (刘鑫)$^{a}$
\quad Tian-wei Bao (鲍天威)$^{a}$
\quad Li Zhang (张力)$^{a}$
\quad Rui-jie Wang (王瑞杰)$^{a}$
\\\quad Jun-guang Lv (吕军光)$^{c}$
\quad Bo-bing Wu (吴伯冰)$^{a}$
\quad Shuang-nan Zhang (张双南)$^{a}$%
}
\maketitle

\address{%
$^a$ Key Laboratory of Particle Astrophysics, Institute of High Energy Physics, CAS, Beijing 100049, China\\
$^b$ University of Chinese Academy of Sciences, Beijing 100049, China\\
$^c$ State Key Laboratory of Particle Detection and Electronics, Institute of High Energy Physics, CAS, Beijing 100049, China\\
}

\begin{abstract}
The High Energy cosmic Radiation Detection (HERD) facility is a space mission designed for detecting cosmic ray (CR) electrons, $\gamma$-rays up to tens of TeV and CR nuclei from proton to iron up to several PeV. The main instrument of HERD is a 3-D imaging calorimeter (CALO) composed of nearly ten thousand cubic LYSO crystals. A large dynamic range of single HERD CALO Cell (HCC) is necessary to achieve HERD's PeV observation objectives, which means that the response of HCC should maintain a good linearity from minimum ionizing particle (MIP) calibration to PeV shower maximum. In order to study the linearity of HCC over such a large energy range, a beam test has been implemented at the E2 and E3 beam lines of BEPC. High intensity pulsed electron beam provided by E2 line are used for producing high energy density within HCC; $\pi^{+}$/proton provided by E3 line are used for HCC calibration. The results show that no saturation effect occurs and the linearity of HCC is better than 10\% from 30 MeV (1 MIP) to 1.1$\times$10$^{3}$ TeV (energy density is 93 TeV/cm$^{3}$), which can meet the requirement mentioned above.
\end{abstract}



\footnotetext[0]{\hspace*{-3mm}\raisebox{0.3ex}{}2016}%
\begin{multicols}{2}

\section{Introduction}
\label{Intro}
HERD is one of the space astronomy payloads of the cosmic light house program onboard the China's Space Station. The main scientific objectives of HERD are indirect dark matter search, precise CR spectrum and composition measurements up to the knee energy and high energy $\gamma$-ray monitoring and survey \cite{bibHerd}. To achieve these scientific objectives, a 3-D imaging and five-side active calorimeter is designed to perform high energy resolution and high statistics measurements of the CR nuclei, electrons and positrons, and $\gamma$-rays in space. HERD CALO consists of 21$\times$21$\times$21 cells corresponding to 55 radiation lengths ($X_{0}$) and 3 nuclear interaction lengths ($\lambda_{I}$) longitudinal, separately. Each HCC is made of a 3$\times$3$\times$3 cm$^{3}$ cubic LYSO crystal and two spiral WLSFs (Saint Gobain BCF91A) as the fluorescence readout. All the WLSF signals are collected by two image intensified CCDs (ICCD) \cite{bibHerd,bibHerdP}. 
\\LYSO (Lutetium Yttrium Orthosilicate with Cerium doping) crystal is one of the best candidates for CALO material. It has the advantages of high density, high light output, short decay time, and application of this crystal in space would be quite easier thanks to the non-hygroscopicity and small temperature coefficient. The general properties of LYSO are described in Ref.~\cite{bibLyso2,bibLyso3,bibLysoCrystal} Theoretically, the fluorescence intensity of LYSO crystal should be linear with the energy deposition induced by incident particles. However, saturation effect may occur at high ionization density of the crystal. The linearity of LYSO crystal light output should be studied and understood before application. Results from Monte Carlo simulation show that a dynamical range of 2$\times$10$^{6}$ is required for HCC \cite{bibHerd}: A minimum detectable signal down to 30 MeV is required for calibration using MIP; meanwhile, the maximum energy deposition induced by PeV proton shower is up to 60 TeV (energy density is about 8 TeV/cm$^{3}$, Fig.~\ref{fProton}). 

\begin{center}
\includegraphics[width=8cm]{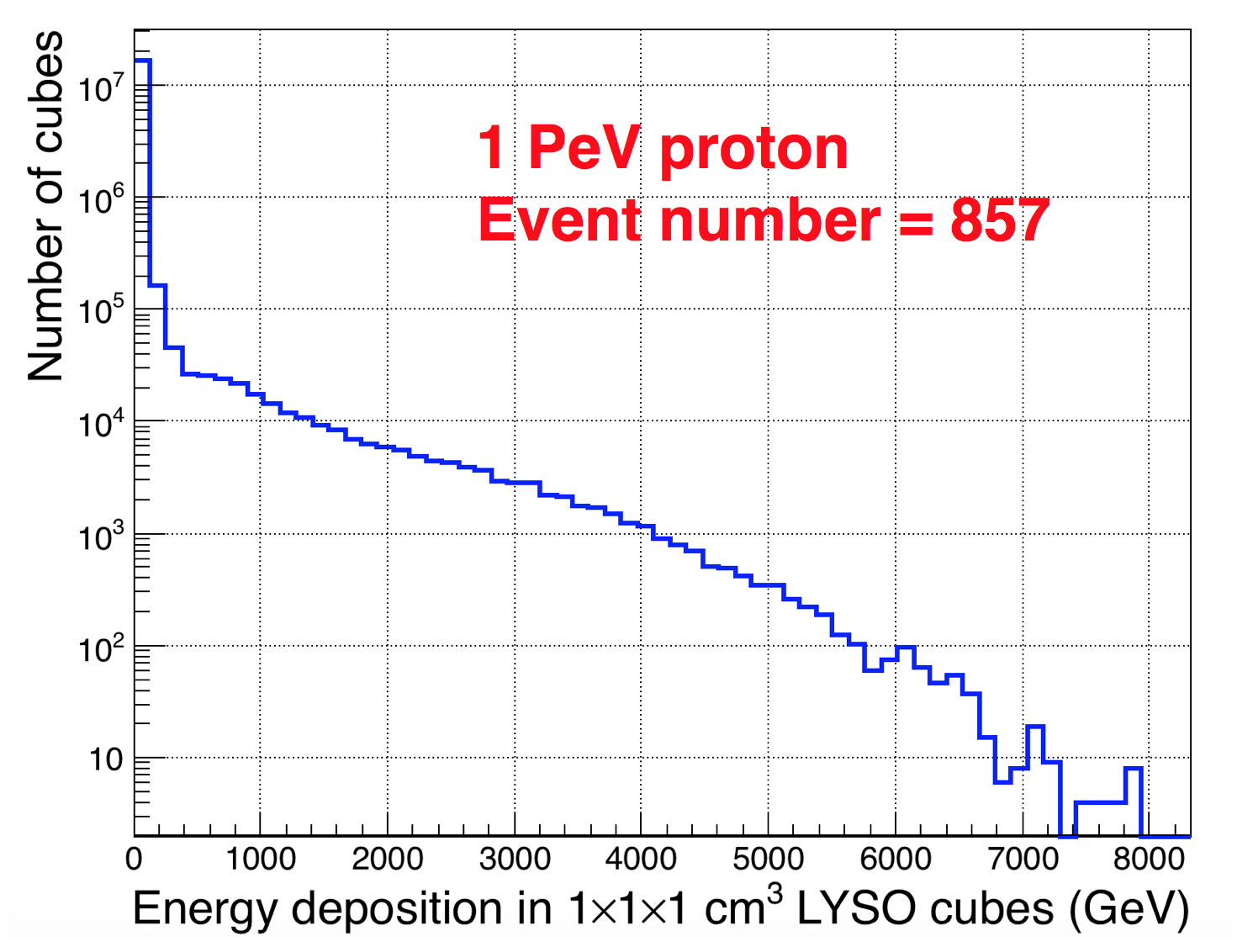}
\figcaption{\label{fProton} Energy deposition in 1$\times$1$\times$1 cm$^{3}$ LYSO cubes with 1 PeV proton incidence. The whole calorimeter consists of 100$\times$100$\times$100 cubes corresponding to 5 $\lambda_{I}$ deep, thus the shower maximum produced by 1 PeV proton can be contained within the calorimeter for those early showered events. The data is derived from Monte Carlo simulation using Fluka 2011 with DPMJET3 \cite{bibFluka}.}
\end{center}

Numerous studies of LYSO linearity have been performed using low energy $\gamma$-rays or electrons, the verified ranges are 511 keV$-$1333 keV \cite{bibLysoLin1}, 65 MeV$-$145 MeV \cite{bibLysoLin2} and 100 MeV$-$1.5 GeV \cite{bibLysoLin3}. There are few calorimeters of space missions designed to measure PeV particles directly, thus little attention has been paid to the linearity of inorganic crystals over a large energy (or energy density) range. 
\\A study of linearity for LSO (Lutetium Orthosilicate with Cerium doping) crystal using high intensity pulsed $\gamma$-rays is described in Ref.~\cite{bibLysoLinHigh}, which demonstrates that the linear response of the crystal ranges from 5.1$\times$10$^{18}$ MeV/(cm$^{2}\cdot$s)$-$1.9$\times$10$^{19}$ MeV/(cm$^{2}\cdot$s). It is reasonable to expect the same linearity of HCC which is made of similar material at that high energy density. However, the verified energy density of this study is too high and it does not cover the required range of HCC.
\\We developed a method of using high intensity pulsed electron beam and $\pi^{+}$/proton beam to measure the linearity of HCC. In this paper, the method and the experimental setup are presented in Section~\ref{EXP}. Data analysis and results are presented in Section~\ref{DA} and the conclusion is briefly shown in Section~\ref{Con}.

\section{Experimental setup}
\label{EXP}
\subsection{Test Beam}
\label{beam}
It is well known that ground accelerators can only produce primary beam particles up to several TeV, thus it is feasible to achieve an energy density up to 8 TeV/cm$^{3}$ by using a bunch of particles passing through HCC and calibrate it by using single beam particle, separately. The measurement is carried out by using the electron beam provided by BEPC E2 Line \cite{bibBEPC}. The beam properties are showed as follows:
\begin{itemize}
\item[\textbullet] Beam energy: 2.5 GeV
\item[\textbullet] Beam intensity: from 10$^{3}$ to 10$^{10}$ electrons/pulse
\item[\textbullet] Beam frequency: 12.5 Hz
\item[\textbullet] Beam size: 4 $\times$ 4 cm$^{2}$
\item[\textbullet] Beam pulse width: 20 ps
\end{itemize}
The collimation size is 1 cm which is smaller than crystal size (3 cm) in order to reduce the geometrical non-uniformities.
Energy deposition of one electron with 2.5 GeV kinetic energy in HCC is about 240 MeV (Section~\ref{E3Result}). HCC is thick enough (2.6$X_{0}$) for 2.5 GeV electrons to produce showers, the part of HCC at the shower center has the highest energy density of about 20 MeV/cm$^{3}$. Thus an adequate beam intensity up to 4$\times$10$^{5}$  electrons/pulse (or energy deposition in HCC up to 96 TeV) is necessary to get a maximum energy density of 8 TeV/cm$^{3}$.
\\BEPC E3 line is a mixed beam of secondary particles includes $\gamma$, e$^{\pm}$, $\pi^{\pm}$, proton and $\mu^{\pm}$. $\pi^{+}$/proton mixed beam with the momentum of 600 MeV/c to 1.2 GeV/c are chosen by beam line magnets for HCC calibration.

\subsection{The HCC readout}
PMTs instead of ICCDs are used during the test to avoid possible non-linearity of the ICCD device. Two WLSFs coupled to a cubic LYSO crystal (from Suzhou Jtcrystal Co., Ltd.) are read out by two PMTs (XP2262) respectively. Neutral density filters (NDF) with different transmittances are attached to PMTs for detecting different energy ranges. The NDFs and the PMT are all sealed in a black plastic cylinder box with a circular aperture of 0.4 mm on top for the WLSF to plug in, as shown in Fig.~\ref{fHCC}. 
\begin{center}
\includegraphics[width=8cm]{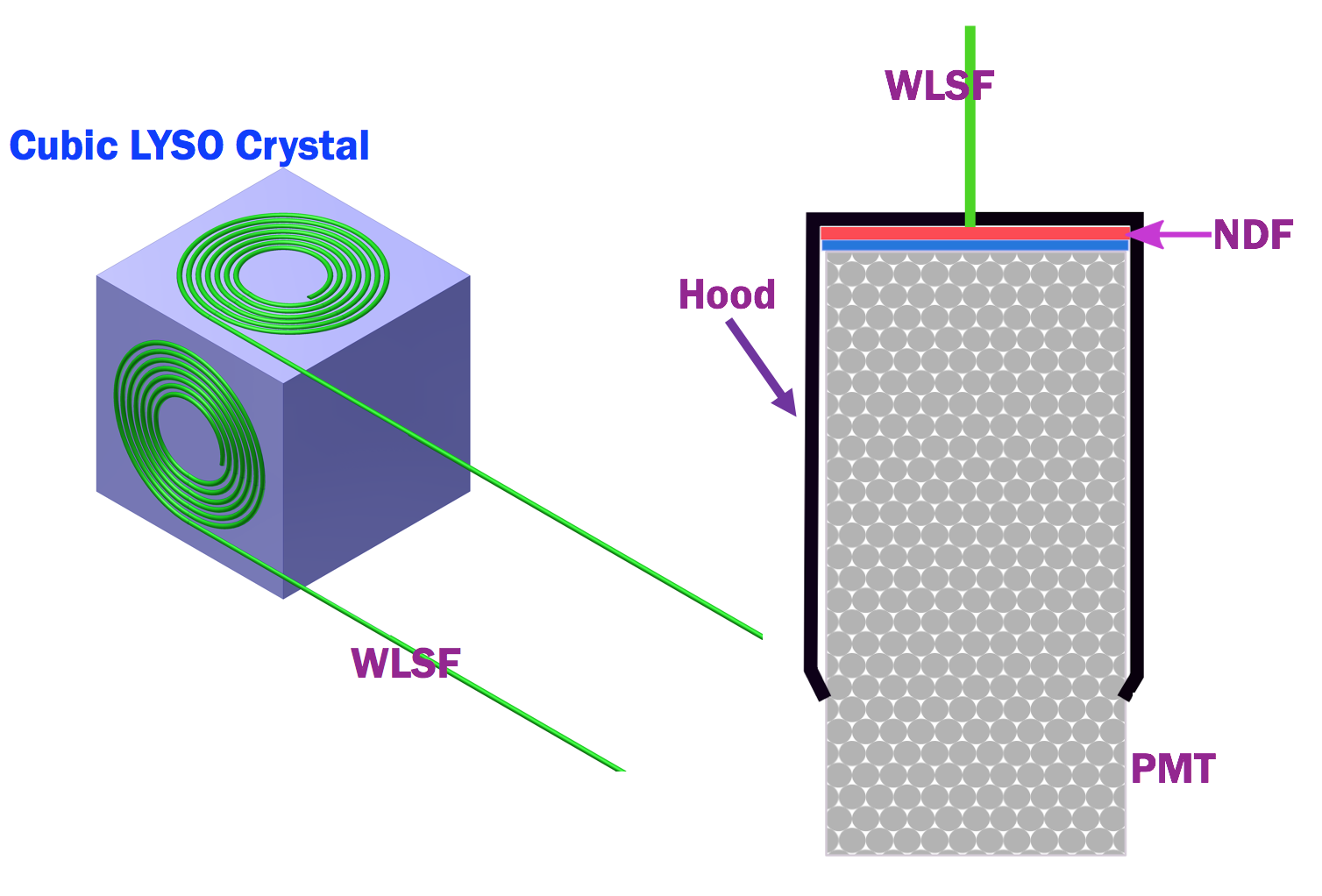}
\figcaption{\label{fHCC} The scheme of HCC (left) and PMT readout (right). }
\end{center}
~\\
~\\
\subsection{Experimental setup at E2 line}
Three detectors are used to calibrate the beam intensity(Fig.~\ref{f2}):
\begin{itemize}
\item[\textbullet] Faraday Cup (FC): FC is the only device which can measure the absolute beam intensity directly. However, FC is not sensitive enough, it is designed to monitor the beam with intensity $>$10$^{6}$ electrons/pulse;
\item[\textbullet] Ionization Chamber (IC):  designed to monitor the beam intensity from 10$^{4}$ to 10$^{7}$ electrons/pulse;
\item[\textbullet] Thin Plastic Scintillator (PS): placed in front of the HCC and attached with 2 PMTs for readout purpose. The effective detecting range is from 10$^{2}$ to 10$^{5}$ electrons/pulse.
\end{itemize}
With cross calibration between the three detectors, a wide beam intensity from 10$^{2}$ to 10$^{9}$ electrons/pulse can then be monitored and precisely measured. 
\begin{center}
\includegraphics[width=8cm]{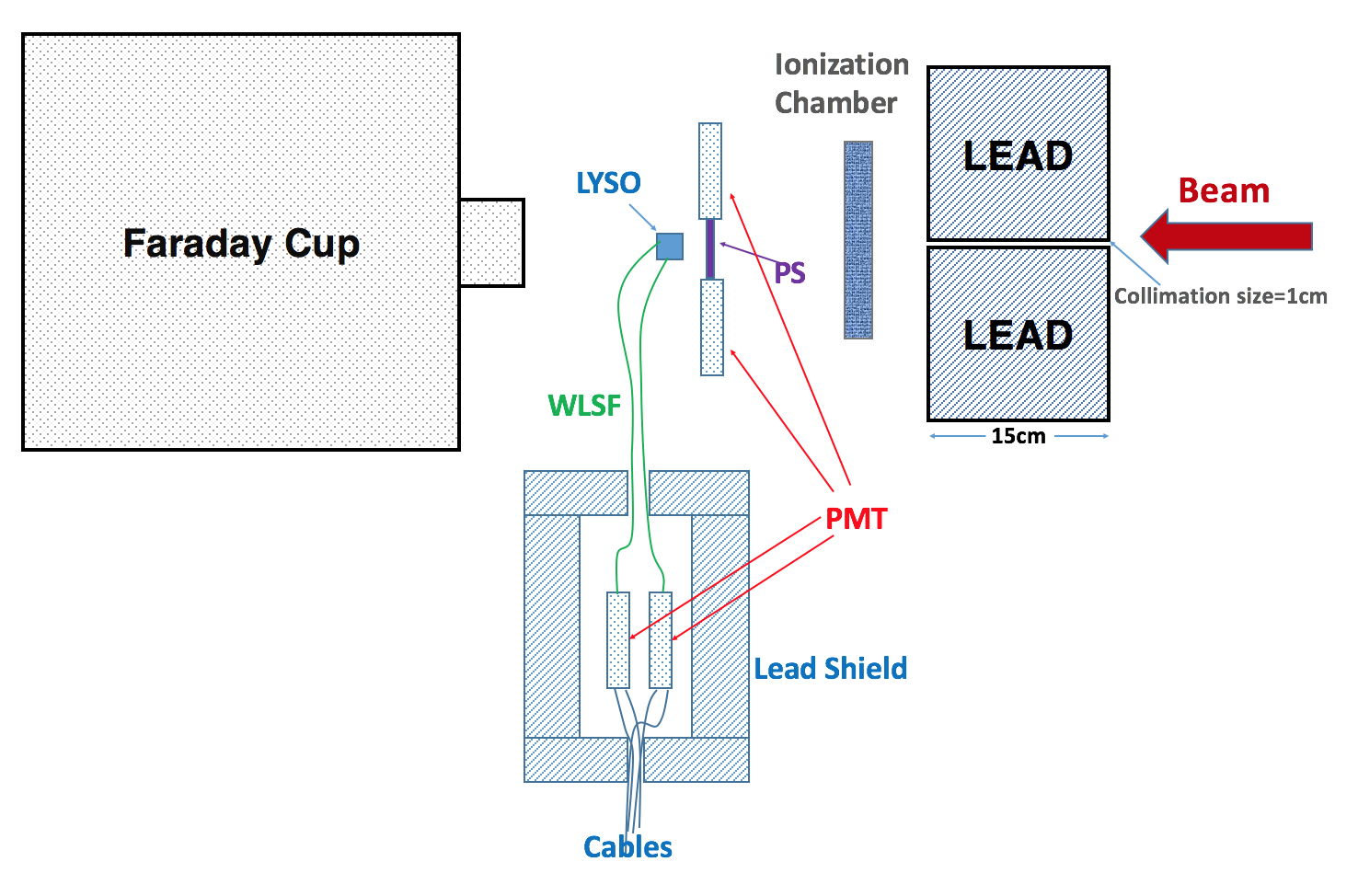}
\figcaption{\label{f2} Schematic diagram of experimental setup for linearity measurement at E2 line.}
\end{center}
A lead wall is built for shielding the detectors from the contamination caused by secondaries from the interaction of beam particles with surrounding materials. Signals from FC and IC are recorded by a 12-bit peak sensing ADC (Mod. V785N), while signals from 4 PMTs are readout by a 4-channel digitizer (Mod. DT5751). Both DAQ modules share a common trigger signal from the accelerator.

\subsection{Experimental setup at E3 line}
\begin{center}
\includegraphics[width=8cm]{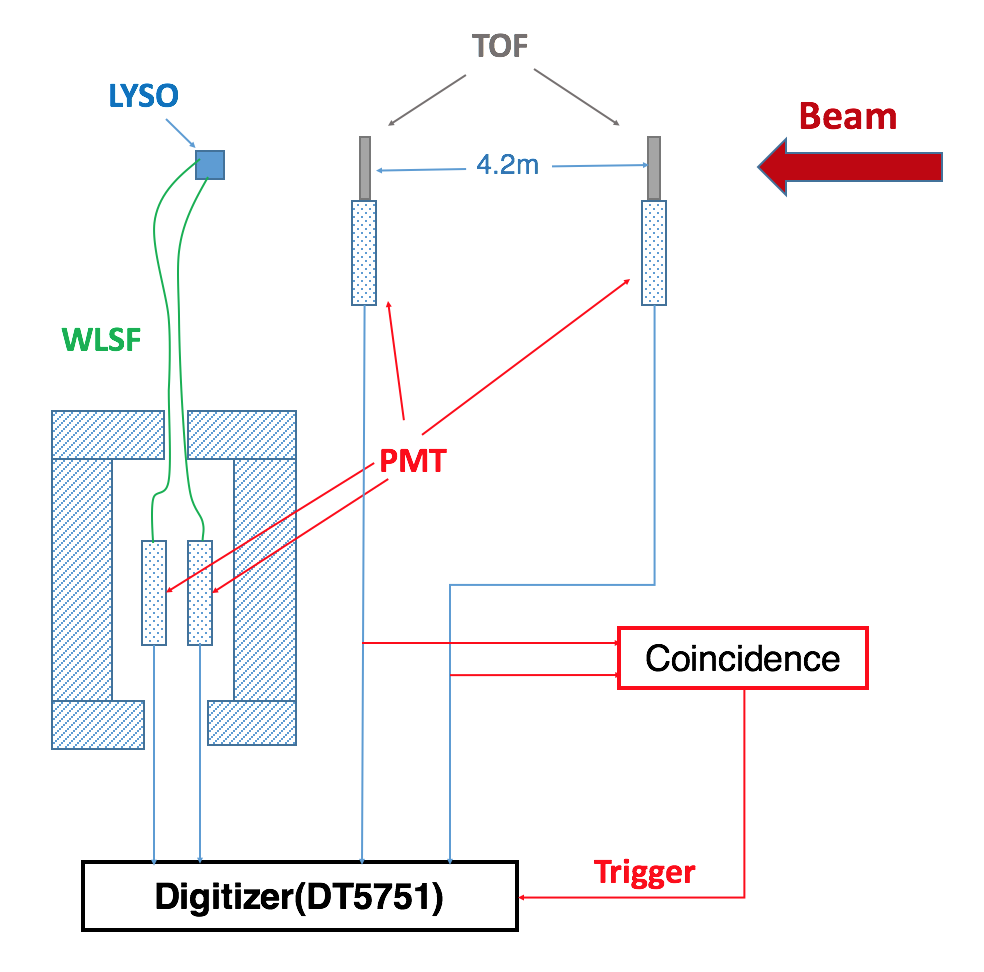}
\figcaption{\label{f3} Schematic diagram of experimental setup for HCC calibration at E3 line, the NDFs are removed and the high voltages for PMTs are raised to detect the MIP signal which is much weaker than electron signal.}
\end{center}

In the configuration at E3 line, a time of flight system (TOF) consists of two plastic scintillators are used for $\pi^{+}$/proton separation as shown in Fig.~\ref{f3}. The trigger is the coincidence of two PMTs attached to the plastic scintillators. DT5751 which can work at a sampling rate of 1 GS/s provides a high time resolution to distinguish proton signals and $\pi^{+}$ signals from TOF (Section~\ref{E3Result}).

\section{Data analysis}
\label{DA}
\subsection{Calibration of beam intensity monitoring devices}
A cross-calibration between FC, IC and PS is performed as shown in Fig.~\ref{f4}.
\begin{center}
\includegraphics[width=8cm]{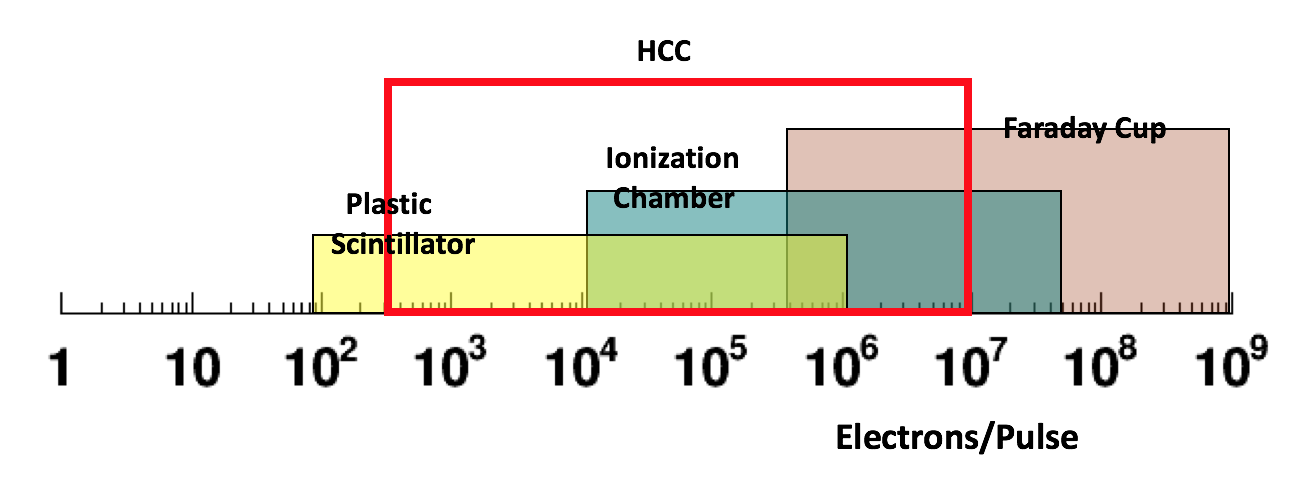}
\figcaption{\label{f4} Principle of beam intensity calibration. Overlaps between the detecting ranges of FC, IC and PS are large enough for a cross-calibration.}
\end{center}

\subsubsection{FC Calibration}

\begin{center}
\includegraphics[width=8cm]{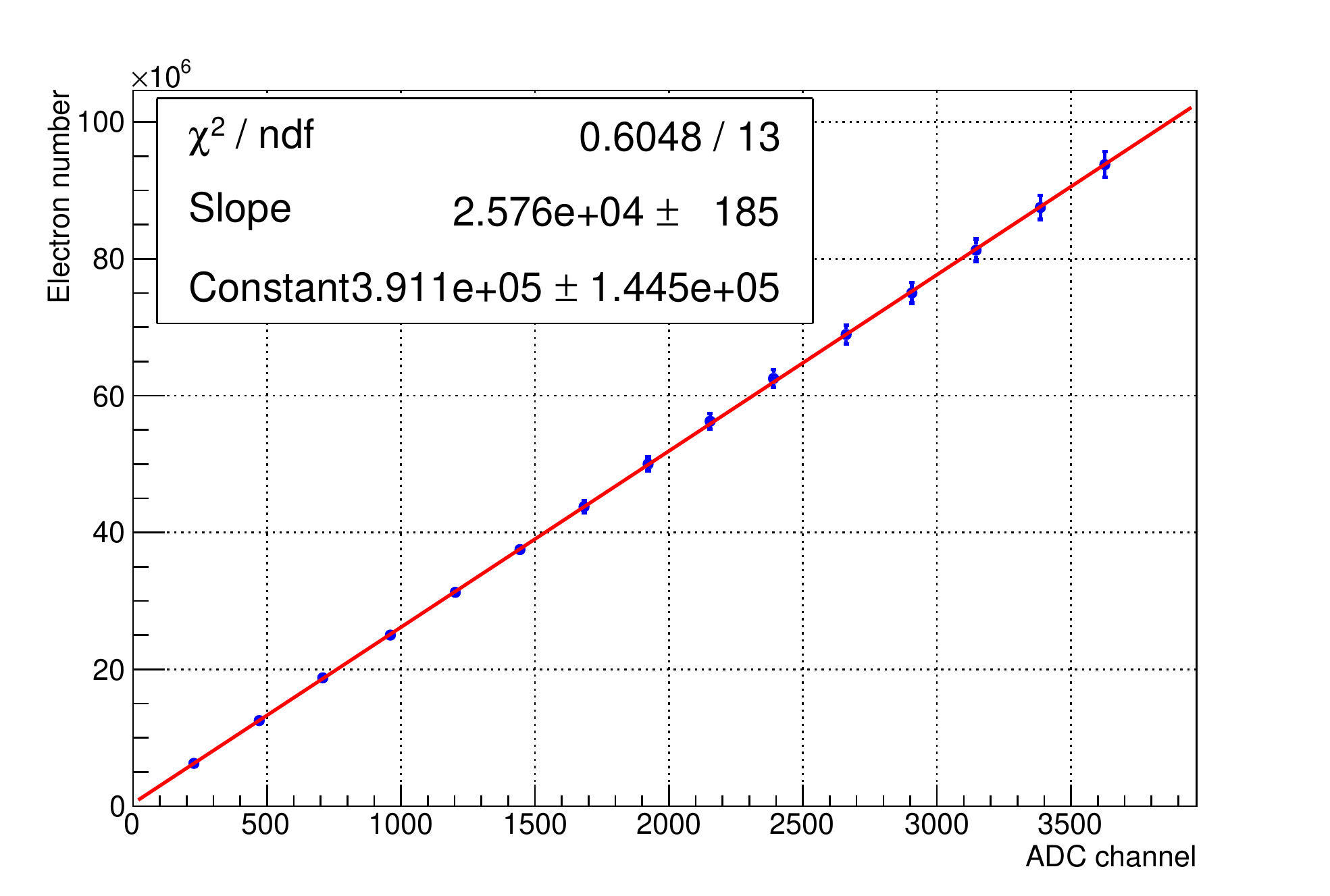}
\figcaption{\label{f5} Relationship between injected electron number and FC output, the x axis is ADC channel; the y axis is number of electrons.}
\end{center}
Fig.~\ref{f5} shows good linearity in ADC's counting range, the calibration factor is obtained by fitting a straight line to data points:
\begin{eqnarray}
\label{eqeq1}
1\ FC\ Channel = 2.58\times10^{4}\pm185\ electrons
\end{eqnarray}
\subsubsection{IC and PS Calibration}
The IC and PS detectors are calibrated by studying the relationship between FC output and IC output (Fig.~\ref{f6}), IC output and PS output (Fig.~\ref{f7}).
\begin{center}
\includegraphics[width=8cm]{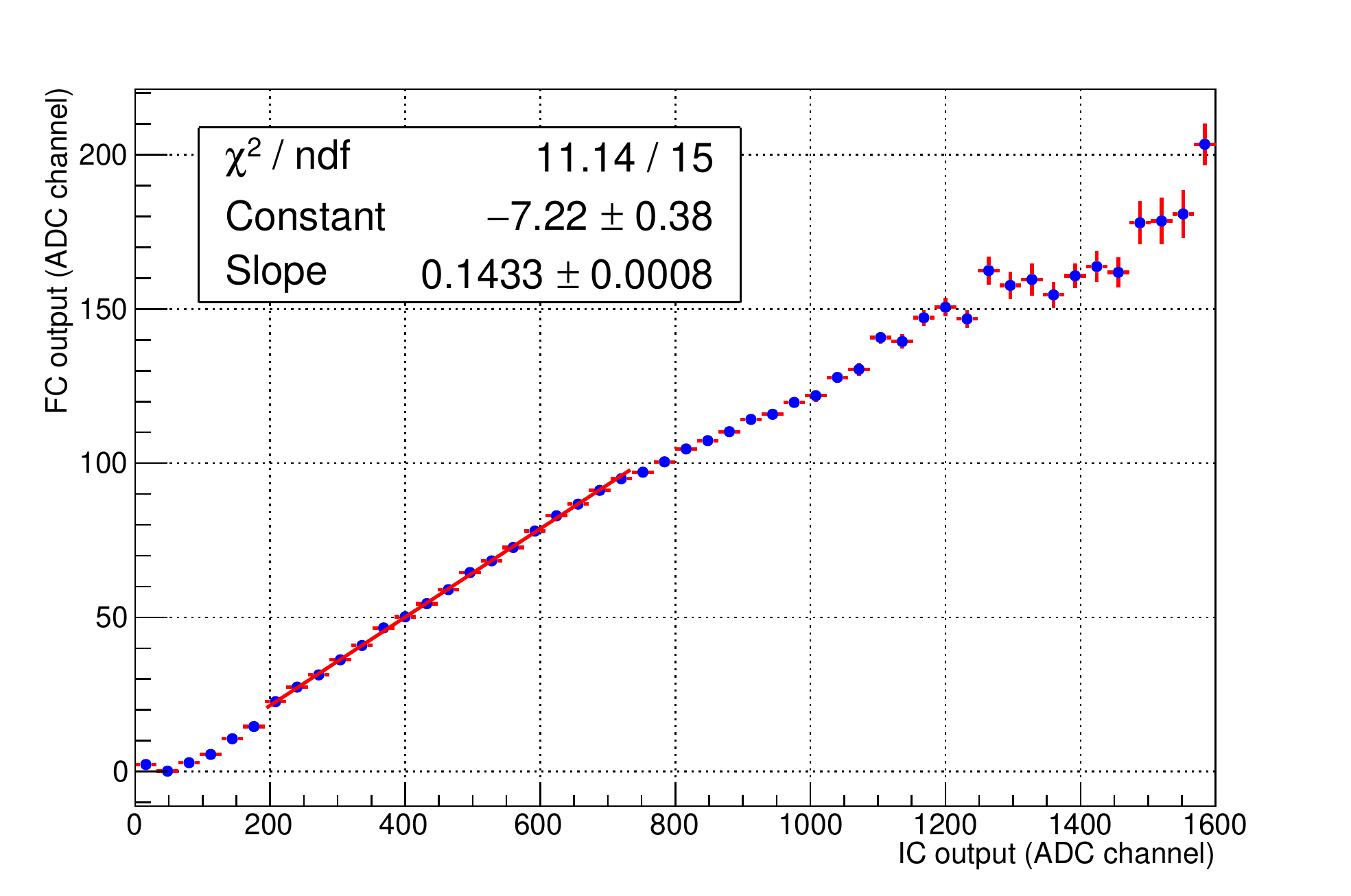}
\figcaption{\label{f6} Profile histogram of IC calibration for displaying the relationship between FC output and IC output.}
\end{center}
ADC channels of IC from 200 to 750 are chosen for fitting, because when ADC(IC)$>$750, rare events would lead large fluctuations, and when ADC(IC)$<$200, signals would be too close to the baseline.
\begin{center}
\includegraphics[width=8cm]{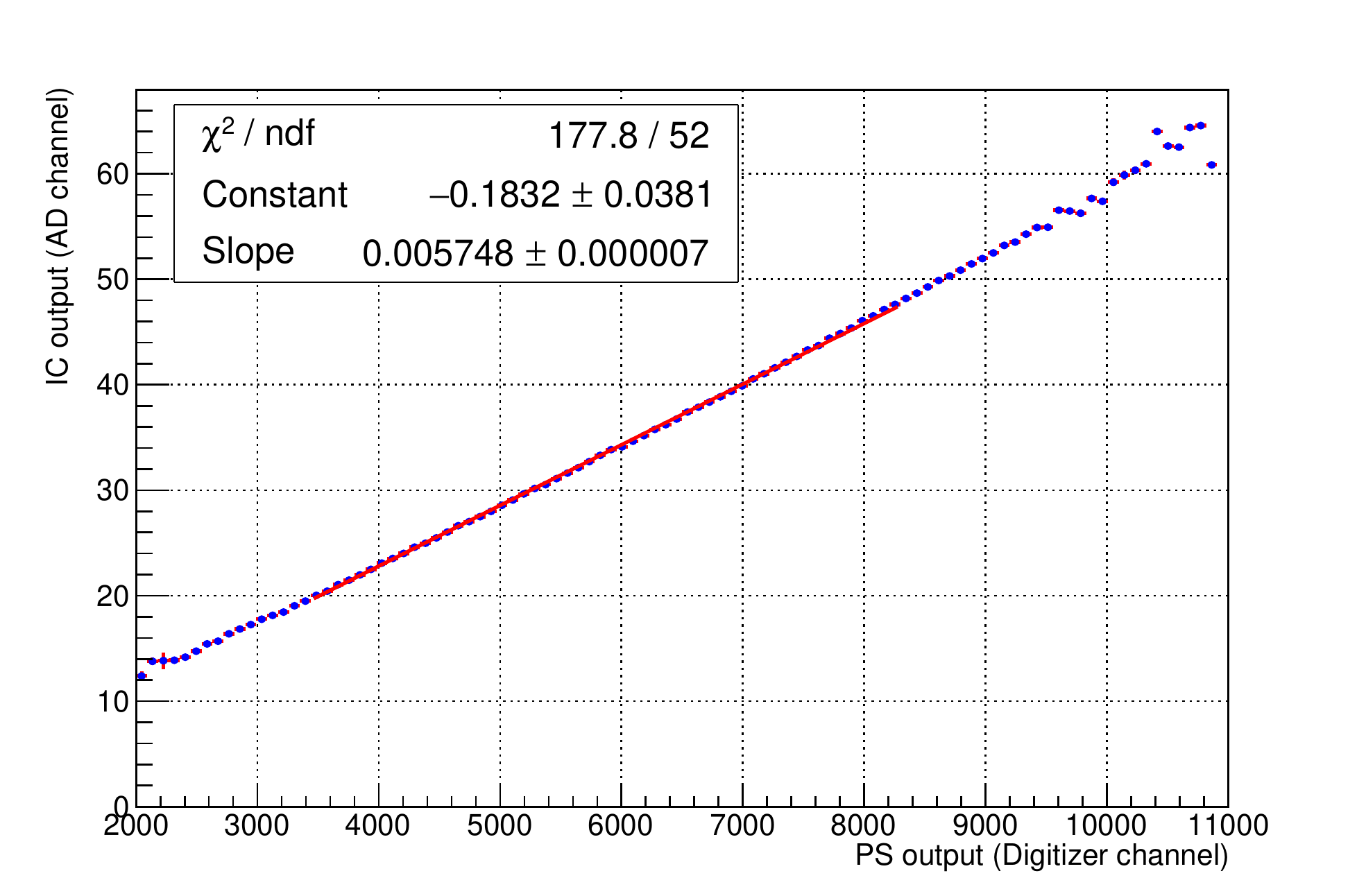}
\figcaption{\label{f7} PS calibration, the x axis represents PS output (Digitizer channel); the y axis is IC output (ADC channel), the histogram shows good linearity.}
\end{center}
The calibration factors of IC and PS are obtained by fitting to the data:
\begin{eqnarray}
\label{eqeq2}
1\ IC\ Channel = 3697\pm26\ electrons
\end{eqnarray}
\begin{eqnarray}
\label{eqeq3}
1\ FC\ Channel = 21.25\pm0.15\ electrons
\end{eqnarray}
\subsection{NDF Calibration}
\label{NDFCali}
A measurement of transmittance of NDFs was performed before the beam test, we used an optical fiber irradiated by a LED as the light source and a PMT for readout. The NDF was installed on a linear stage which could provide high precision linear movement. The original signal and attenuated signal could be obtained by moving the NDF to the position between the light source and the PMT. Fig.~\ref{ndf} shows the calibration result of one NDF.
\begin{center}
\includegraphics[width=8cm]{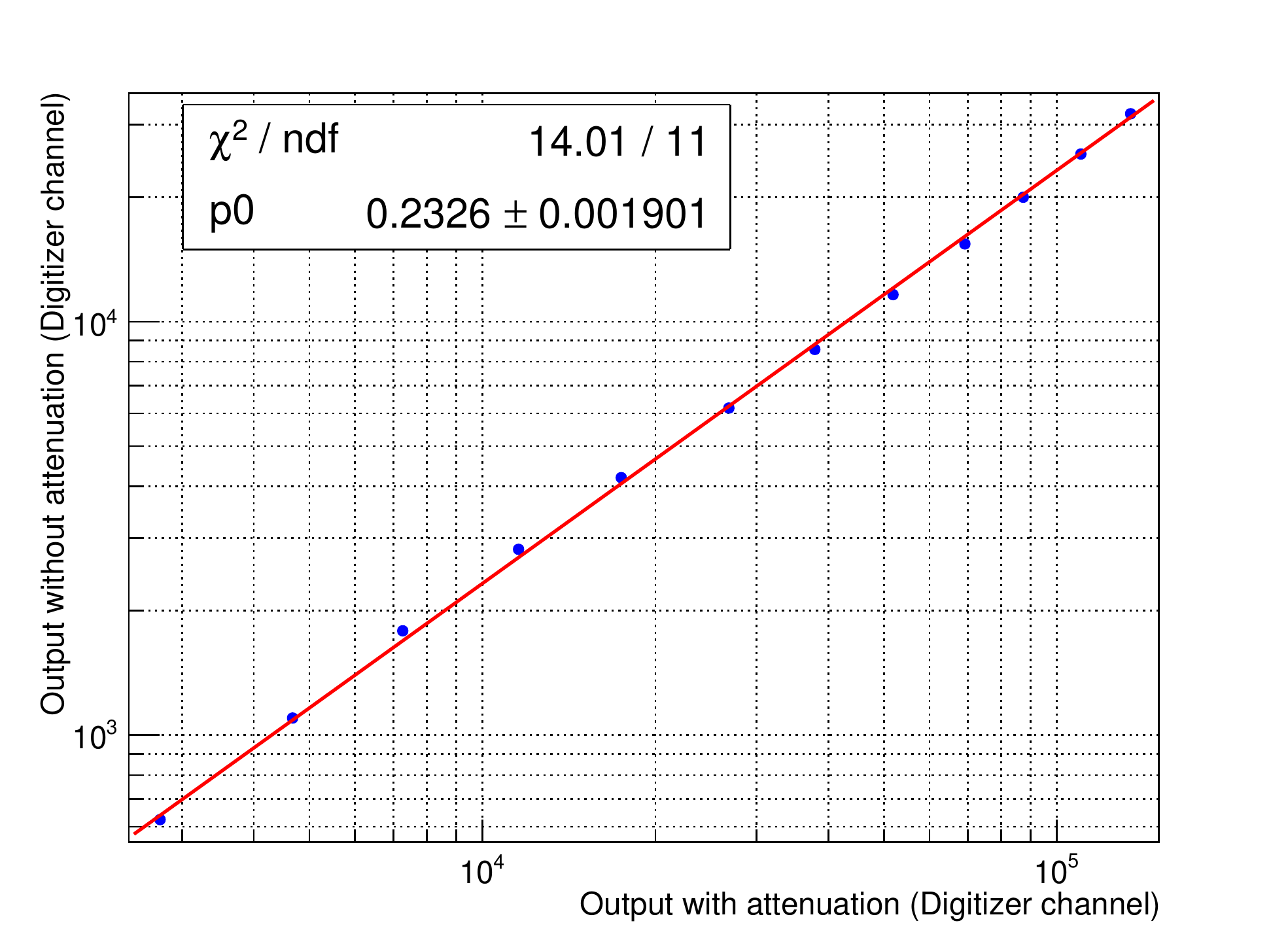}
\figcaption{\label{ndf} Calibration result of one NDF with 0.6 optical density(O.D., $Transmittance=10^{-O.D.}$).}
\end{center}

\end{multicols}

\begin{table}
\tabcaption{ \label{t1}  Calibration results of the NDFs used in the test.}
\footnotesize
\begin{tabular*}{170mm}{@{\extracolsep{\fill}}ccccccc}
\toprule No. & Attached PMT  & Diameter (mm) & Optical density & Nominal transmittance &  Calibrated transmittance \\
\hline
1 & L & 504 & 0.6 & 0.251 & 0.2326 \\
2 & L & 504 & 2 & 0.01 & 0.0118 \\
3 & H & 504 & 1 & 0.1 & 0.0828 \\
4 & H & 504 & 3 & 0.001 & 0.0010 \\
\bottomrule
\end{tabular*}%
\end{table}

\begin{multicols}{2}

The total transmittance of attached NDFs are 2.8$\times$10$^{-3}$ and 8.3$\times$10$^{-5}$ (Table~\ref{t1}) for the two PMTs (defined as PMT\_L, PMT\_H), thus the intensity of light reached to the windows of PMT\_L is about 34 times higher than that reached to PMT\_H. The dynamic range of the HCC readout system is then extended naturally by combining the two PMTs.

\subsection{HCC Calibration using E3 beam}
\label{E3Result}
Most of E3 beam particles are protons and $\pi^{+}$, TOF data which clearly illustrates the difference between them in Fig.~\ref{f9} is needed to separate $\pi^{+}$ from proton to get a clean spectrum as shown in Fig.~\ref{f10}. 

\begin{center}
\includegraphics[width=8cm]{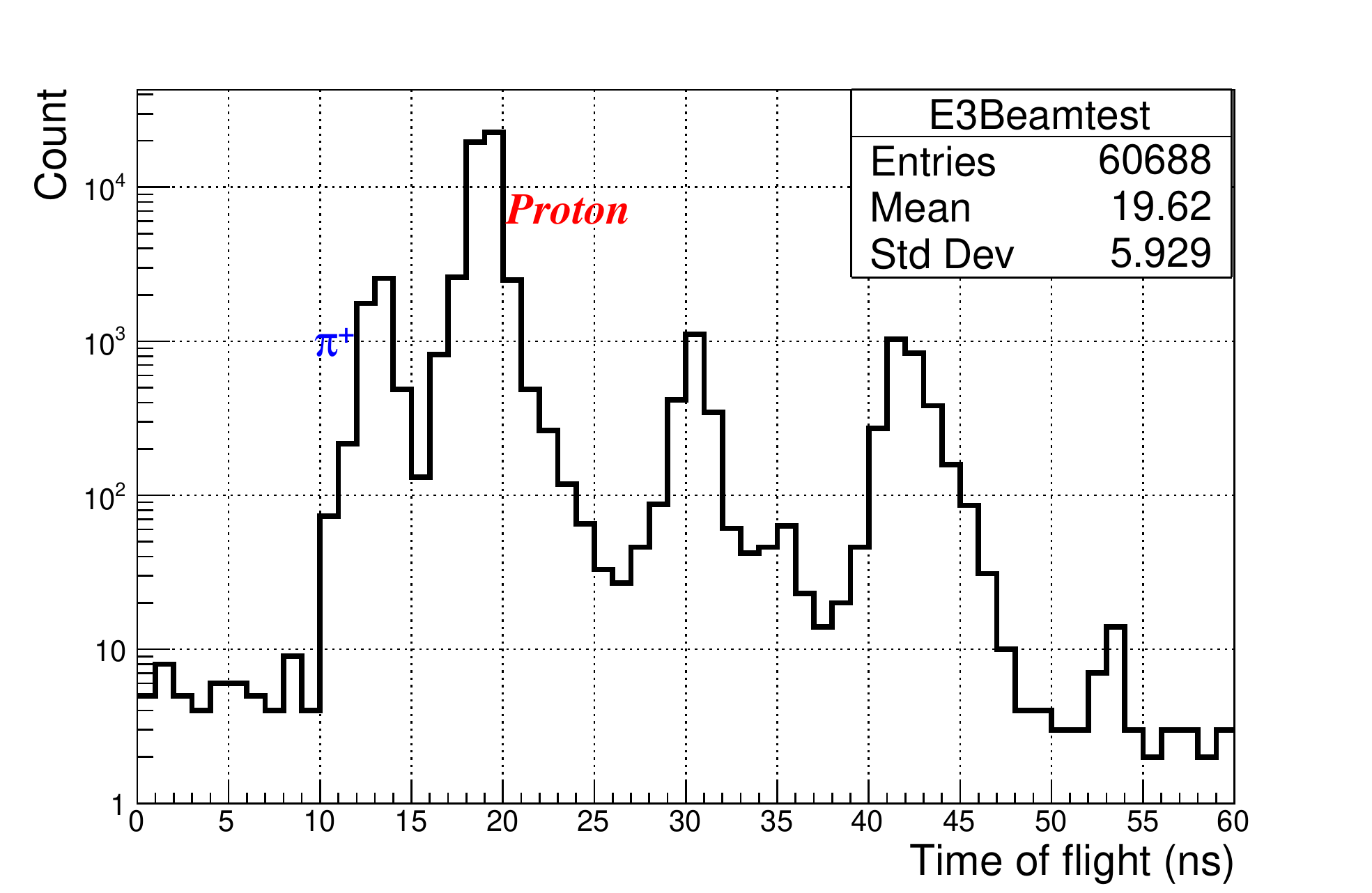}
\figcaption{\label{f9} Time of flight distribution measured by TOF system for E3 beam at 800 MeV/c. The unmarked peaks are caused by accidental coincidences.}
\end{center}

\begin{center}
\includegraphics[width=8cm]{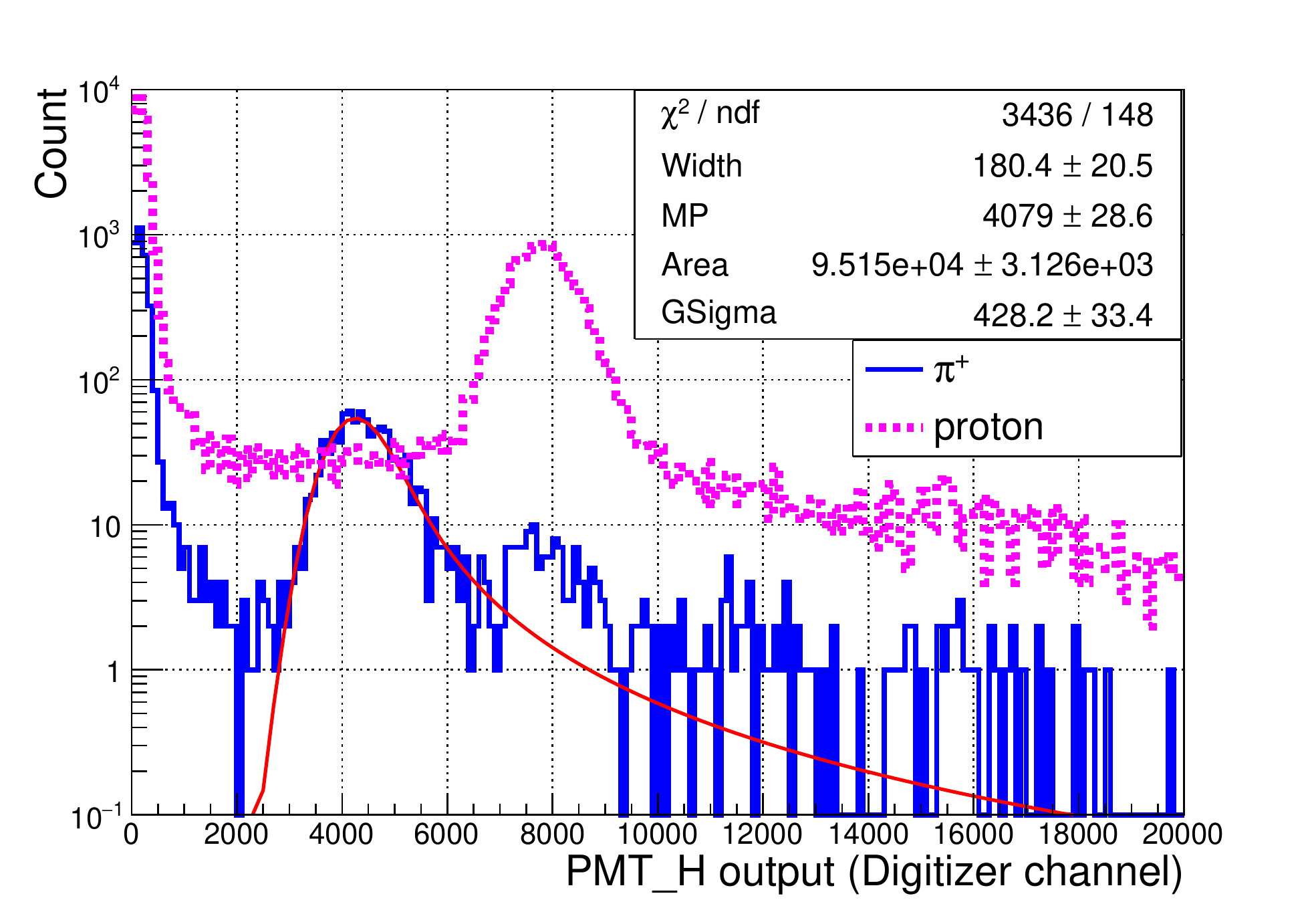}
\figcaption{\label{f10} Use the TOF data to separate $\pi^{+}$ from protons and use landau convoluted Gaussian function to fit $\pi^{+}$ peak. The incident particle is identified as a  $\pi^{+}$ when 10 ns$<$TOF$<$15 ns or a proton when 16 ns$<$TOF$<$21 ns.}
\end{center}

Energy deposition in HCC for different particles is then calculated by Monte Carlo simulation using Geant 4.9.6~\cite{bibGeant4}. The simulation environment is built based on the experimental setup. The results are shown in Fig.~\ref{f11} and Table~\ref{t2}.

Energy deposition is nearly constant for $\pi^{+}$ in the case of normal incident when its momentum $>$400 MeV/c; for 2.5 GeV electrons, energy deposition is much higher because electromagnetic showers are initiated inside the crystal.
\begin{center}
\includegraphics[width=8cm]{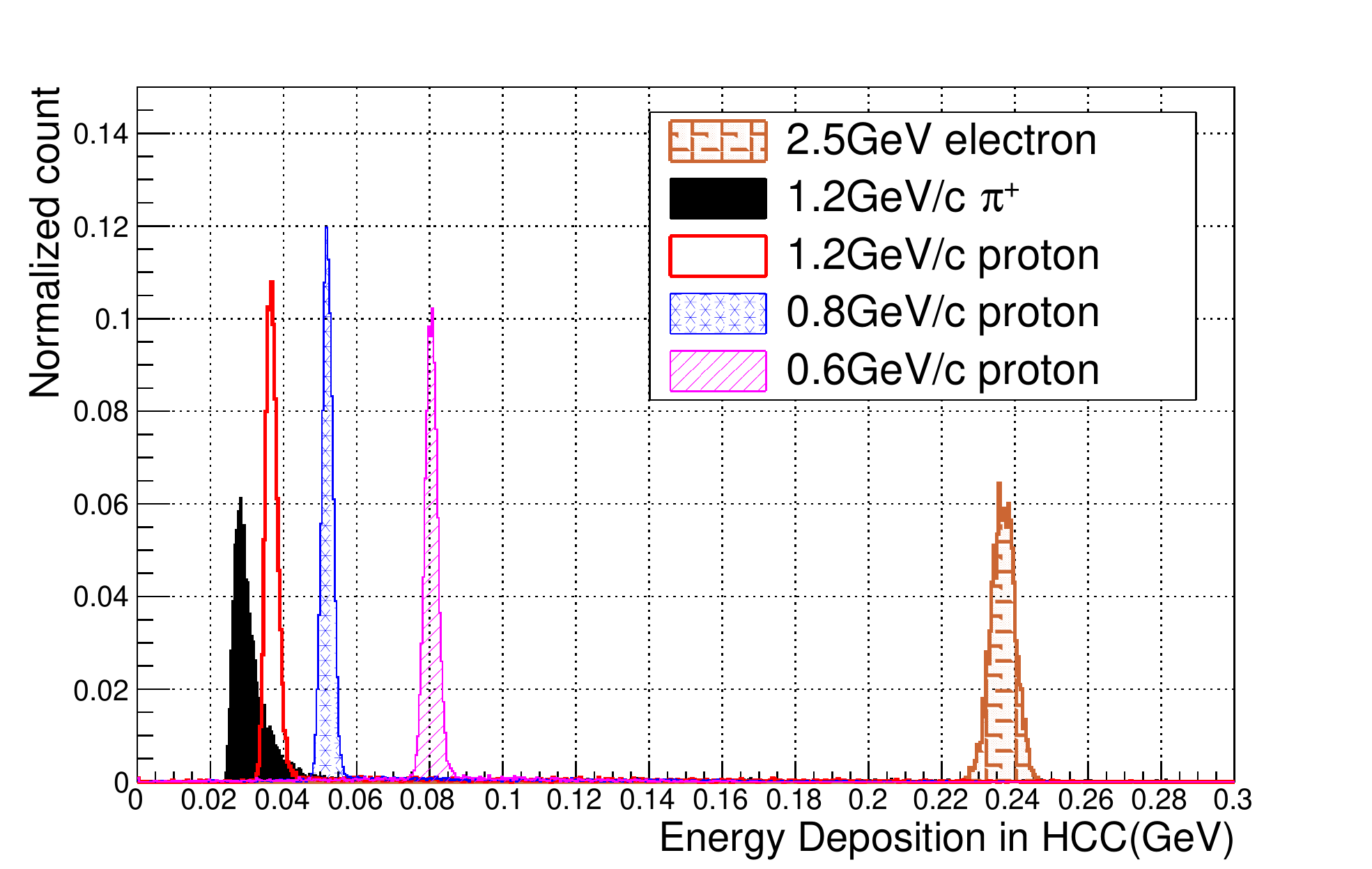}
\figcaption{\label{f11} Energy deposition in HCC for $\pi^{+}$, protons and electrons derived from Mote Carlo simulation data.}
\end{center}

\subsection{Results and Discussions}
\label{Result}
\begin{figure*}[htbp]
\begin{subfigure}{0.5\textwidth}
\centering
\includegraphics[width=0.85\textwidth]{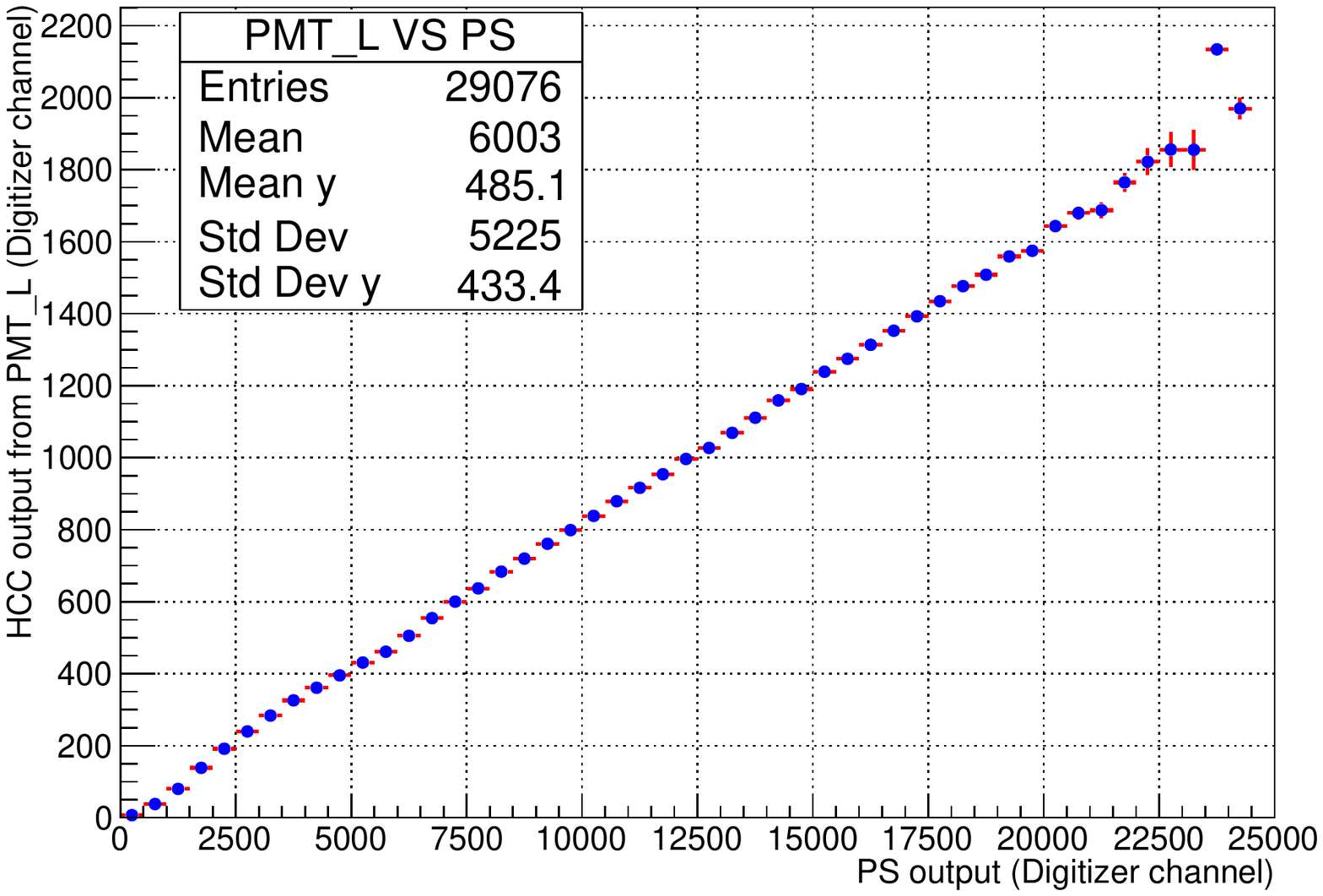}
\caption*{(a)}
\end{subfigure}
\begin{subfigure}{0.5\textwidth}
\centering
\includegraphics[width=0.85\textwidth]{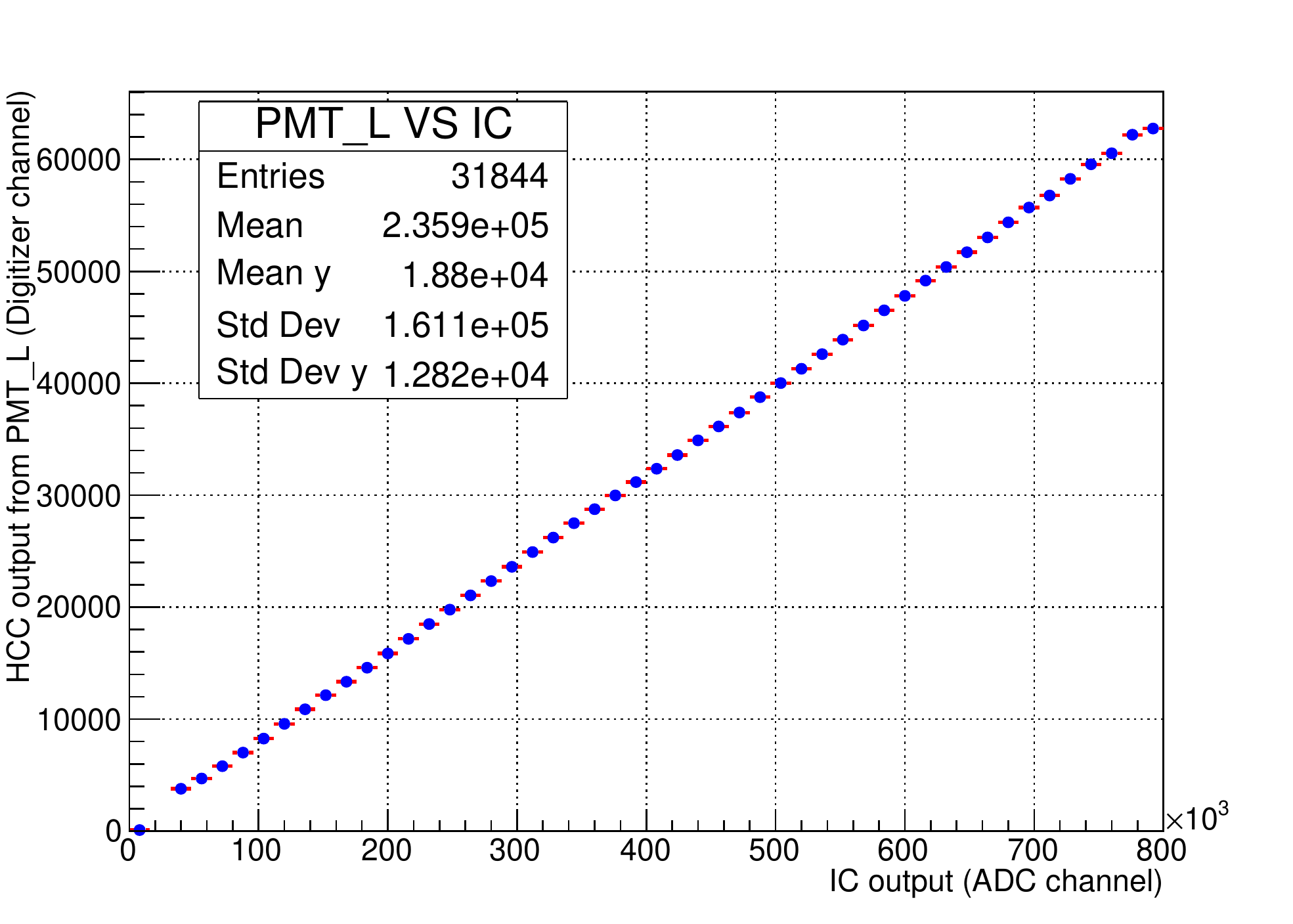}
\caption*{(b)}
\end{subfigure}

\begin{subfigure}{0.5\textwidth}
\centering
\includegraphics[width=0.85\textwidth]{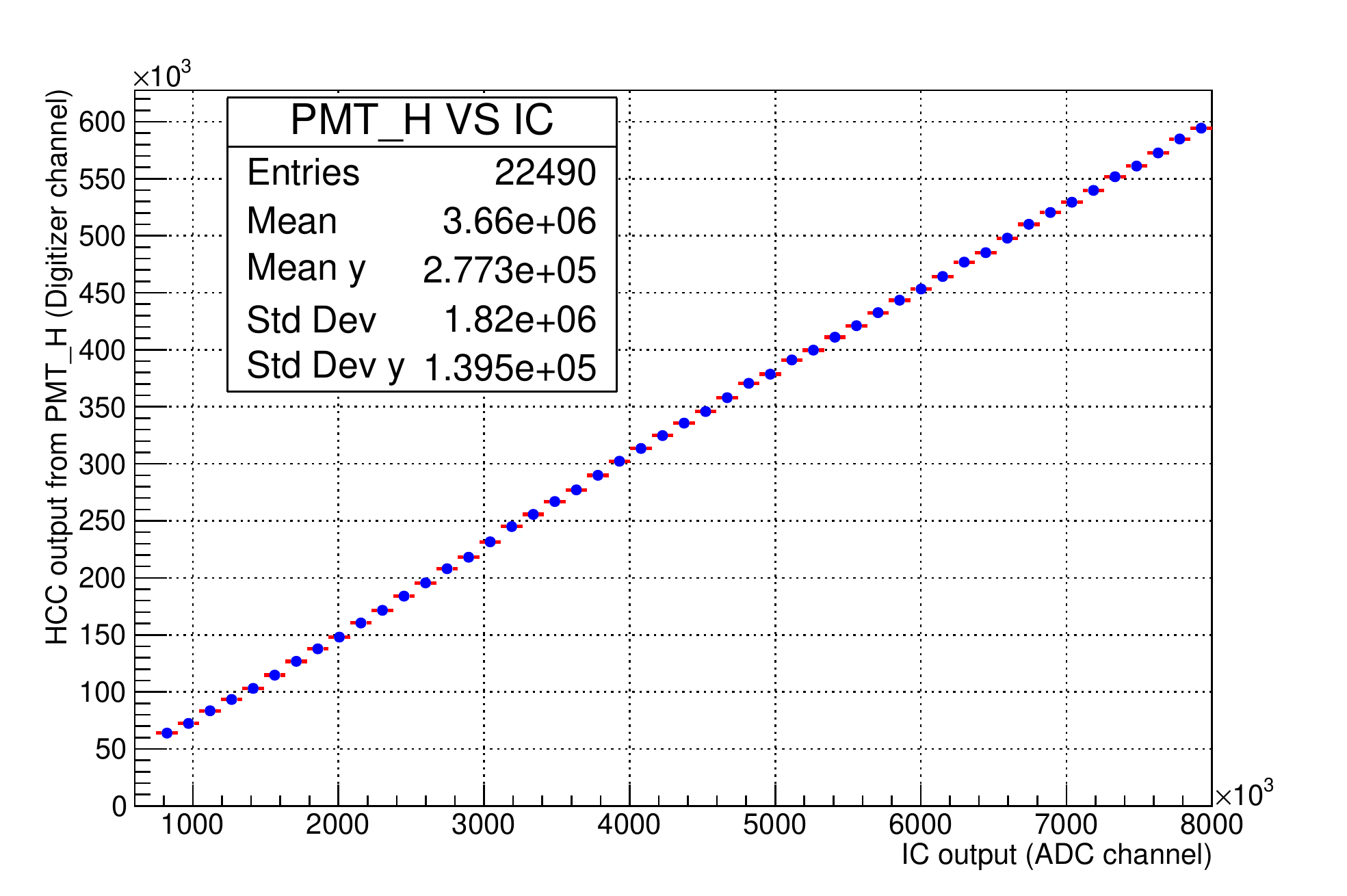}
\caption*{(c)}

\end{subfigure}
\begin{subfigure}{0.5\textwidth}
\centering
\includegraphics[width=0.85\textwidth]{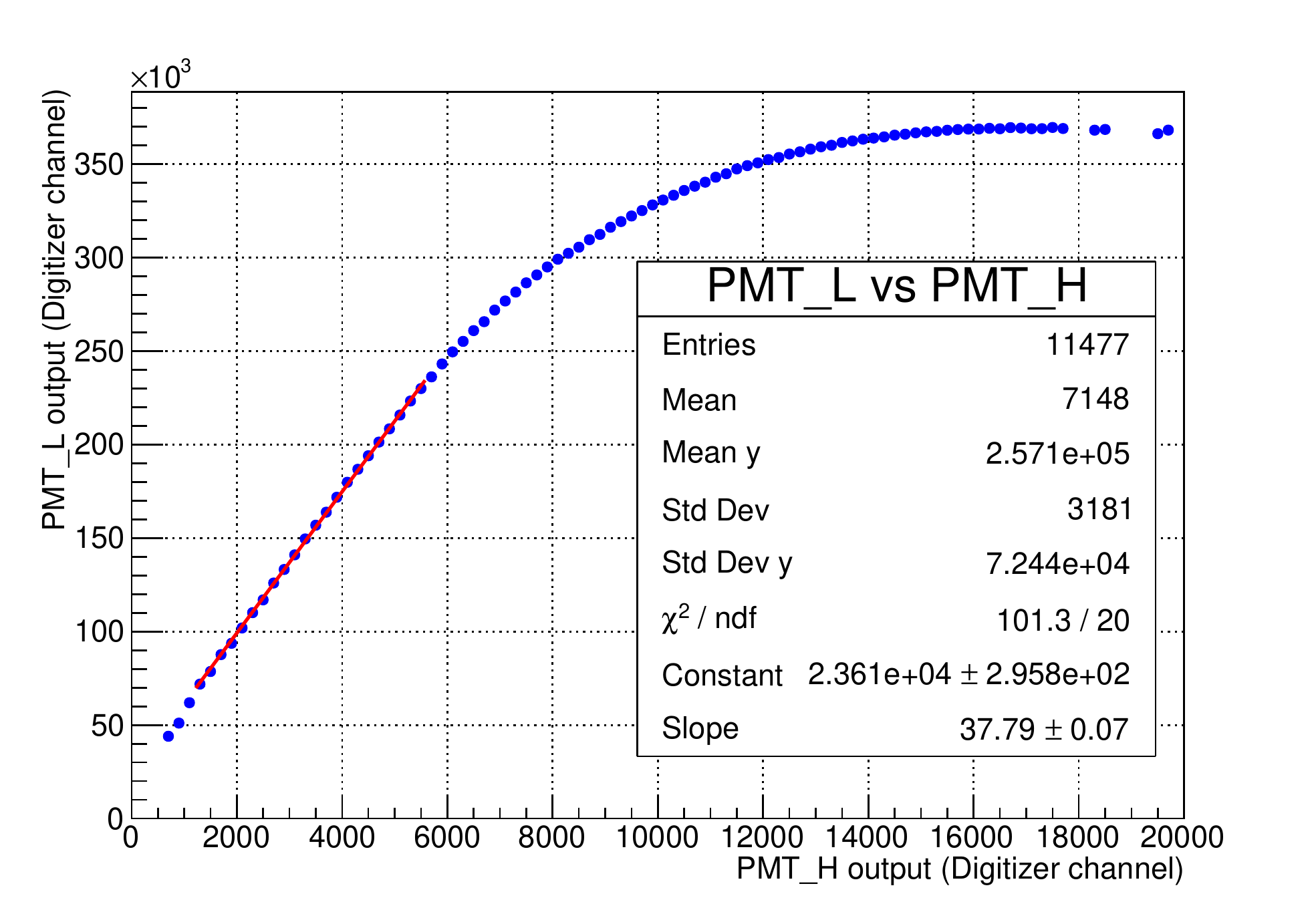}
\caption*{(d)}
\end{subfigure}
\caption{Linearity of HCC at different beam intensities (a,b,c) and correlation between output of PMT\_L and PMT\_H (d). }
\label{f12}
\end{figure*}

The results of linearity at different beam intensities as well as correlation between output of PMT\_L and PMT\_H are illustrated in Fig.~\ref{f12}. 
\\To calculate the equivalent HCC output for proton and $\pi^{+}$, the NDFs' attenuate effect and the change of high voltage for PMTs should be taken into account:

\begin{equation}
\label{eqeq2}
HCC_{\pi^{+},proton} = A_{HV}A_{F}S_{\pi^{+},proton}
\end{equation}

\begin{figure}[H]
\begin{subfigure}{0.5\textwidth}
\centering
\includegraphics[width=0.9\textwidth]{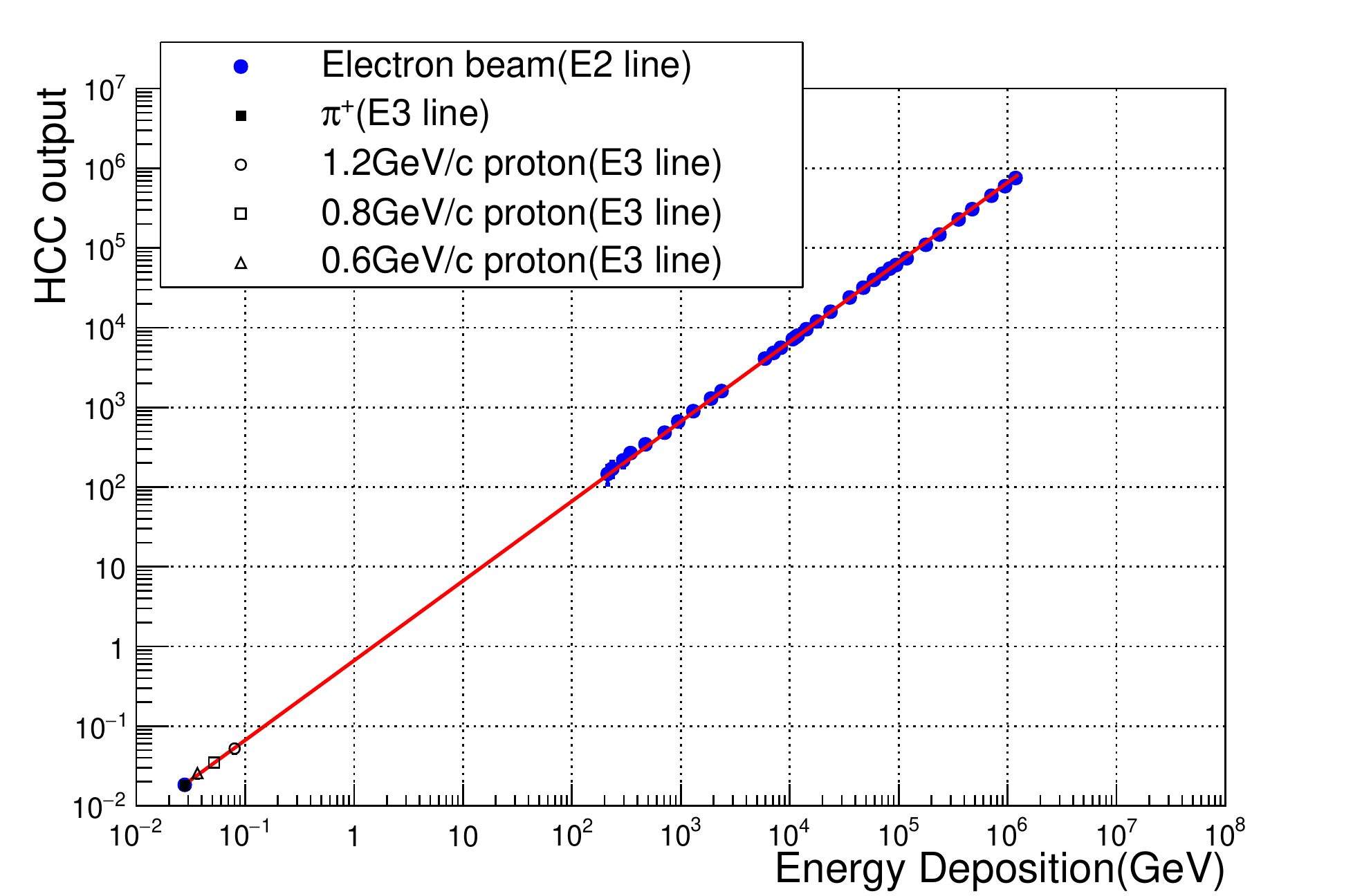}
\caption{}
\label{f14_1}
\end{subfigure}
\begin{subfigure}{0.5\textwidth}
\centering
\includegraphics[width=0.9\textwidth]{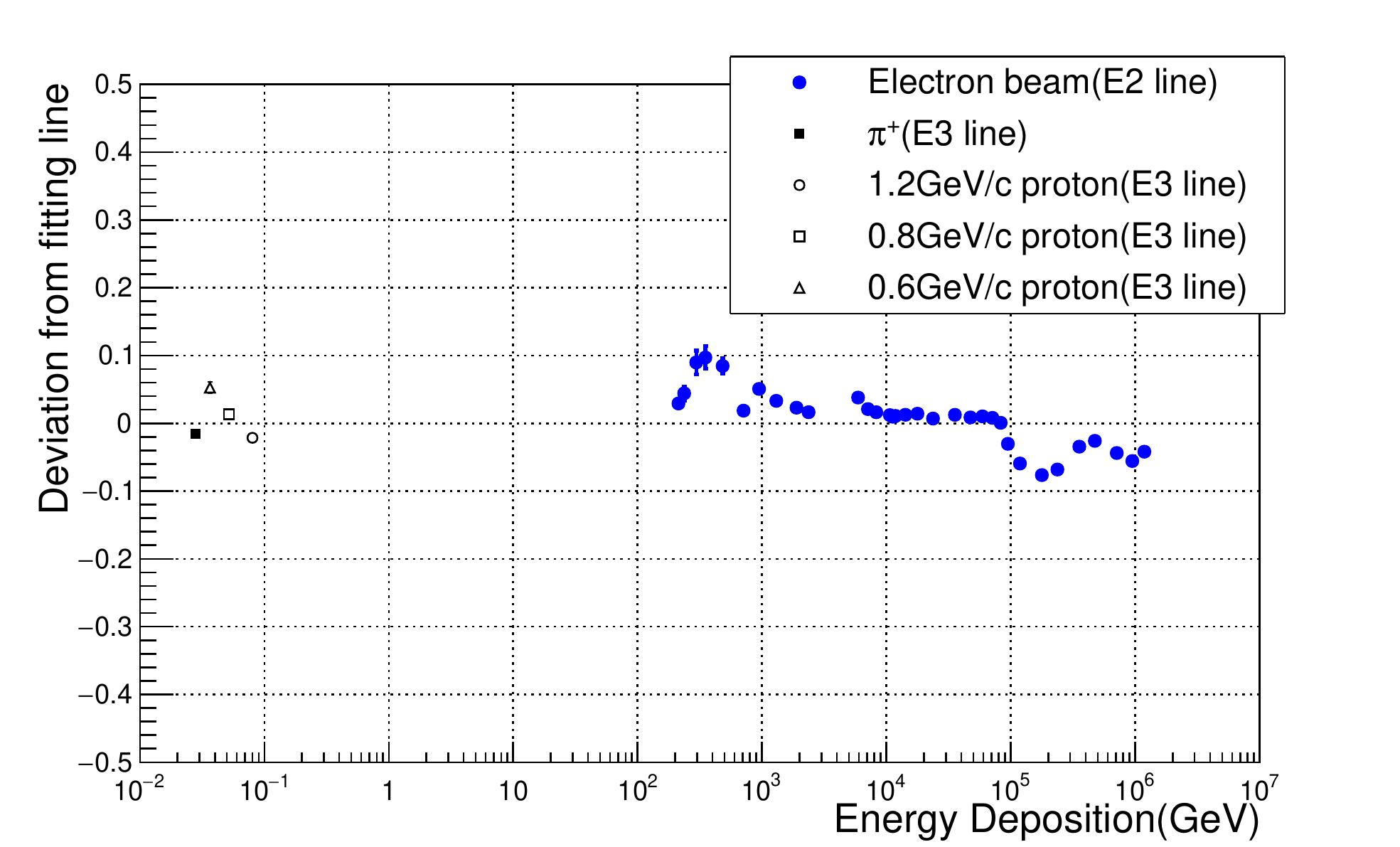}
\caption{}
\label{f14_2}
\end{subfigure}
\caption{Linearity of HCC (a) and deviations from the fitting line (b) for all ranges. The magnets at E2 cannot be precisely controlled to achieve an intensity lower than 200 electrons/pulse, thus a huge gap without measurement points is located in between electrons and $\pi^{+}$/protons.}
\label{f14}
\end{figure}

Where $A_{HV}$ represents the high voltage correction factor which is defined as the ratio of two gains of PMTs used by E2 test and E3 test;  $A_{F}$ represents the total transmittance of NDFs used in E2 test;  $S_{\pi^{+},proton}$ represents the PMT output given by E3 test (Table~\ref{t2}).
Fig.~\ref{f14} shows the profile histogram which combines all the results for different intensity ranges as well as $\pi^{+}$ and protons.

\end{multicols}
\begin{table}
\tabcaption{ \label{t2}  Summary of simulation results and experimental results of beam test at E3.}
\footnotesize
\begin{tabular*}{170mm}{@{\extracolsep{\fill}}ccccccc}
\toprule 
Particle type & Momentum (MeV/c) & \tabincell {c} {Energy deposition from  \\ MC simulation data(MeV)} & \tabincell {c} {Peak Value from experimental \\data
(Digitizer channel) }& \tabincell {c} {Sigma from experimental\\ data
(Digitizer channel)} \\
\hline
$\pi^{+}$ & 600-1200 & 28.0 & 4079 & 428.2\\
Proton & 600 & 80.4 & 11644.6 & 783.1  \\
Proton & 800 & 52.0 & 7796.4 & 630.9  \\
Proton & 1200 & 36.6 & 5715.4& 564.8 \\
Electron & $E_{kinetic}$= 2.5GeV& 236.7 &  &  \\
\bottomrule
\end{tabular*}%
\end{table}

\begin{multicols}{2}

The maximum deviation is defined as linearity of HCC. Fig.~\ref{f14} shows the linearity is better than 10\% when energy deposition (energy density) in HCC are from 205 GeV (17 GeV/cm$^{3}$ ) to 1.1$\times$10$^{3}$ TeV (93 TeV/cm$^{3}$) which covers the requirement. The $\pi^{+}$ and proton signals agree well with the fitting function.

There are some factors that may affect the measurement result of linearity:
\begin{itemize}
\item[a)] Non-linearity of PMTs: Non-linearity of two PMTs was studied before the beam test. The details of measurement are described in \cite{bibPMT}. Result shows the non-linearity of PMT is less than 2\% when the output signal is not higher than 2 V. NDFs with high attenuation are used in the beam test to ensure the output signals are lower than 2 V. 
\item[b)] Stability of PMTs: Before data acquisition, the high voltage of PMTs are usually switched on at least two hours in advance, the response of PMTs gradually become stable in this period. The temperature variation may also cause a drift. To minimize such effect, one full data acquisition is completed within 3 hours and the data acquisition is repeated several times. The temperature variation during the beam test is smaller than 2\textcelsius. Therefore, this residual influence is negligible comparing to the calibration error.
\item[c)] Backgrounds at E2 line: The backgrounds are secondary particles from the interaction of beam electrons with shielding material. Most of the backgrounds are X-rays and $\gamma$-rays. One part of them are absorbed by the LYSO crystal. But the ratio of signal to background is lower than 2\% (Fig.~\ref{f15}) which do not have significant impacts on the measurement.

\begin{center}
\includegraphics[width=8cm]{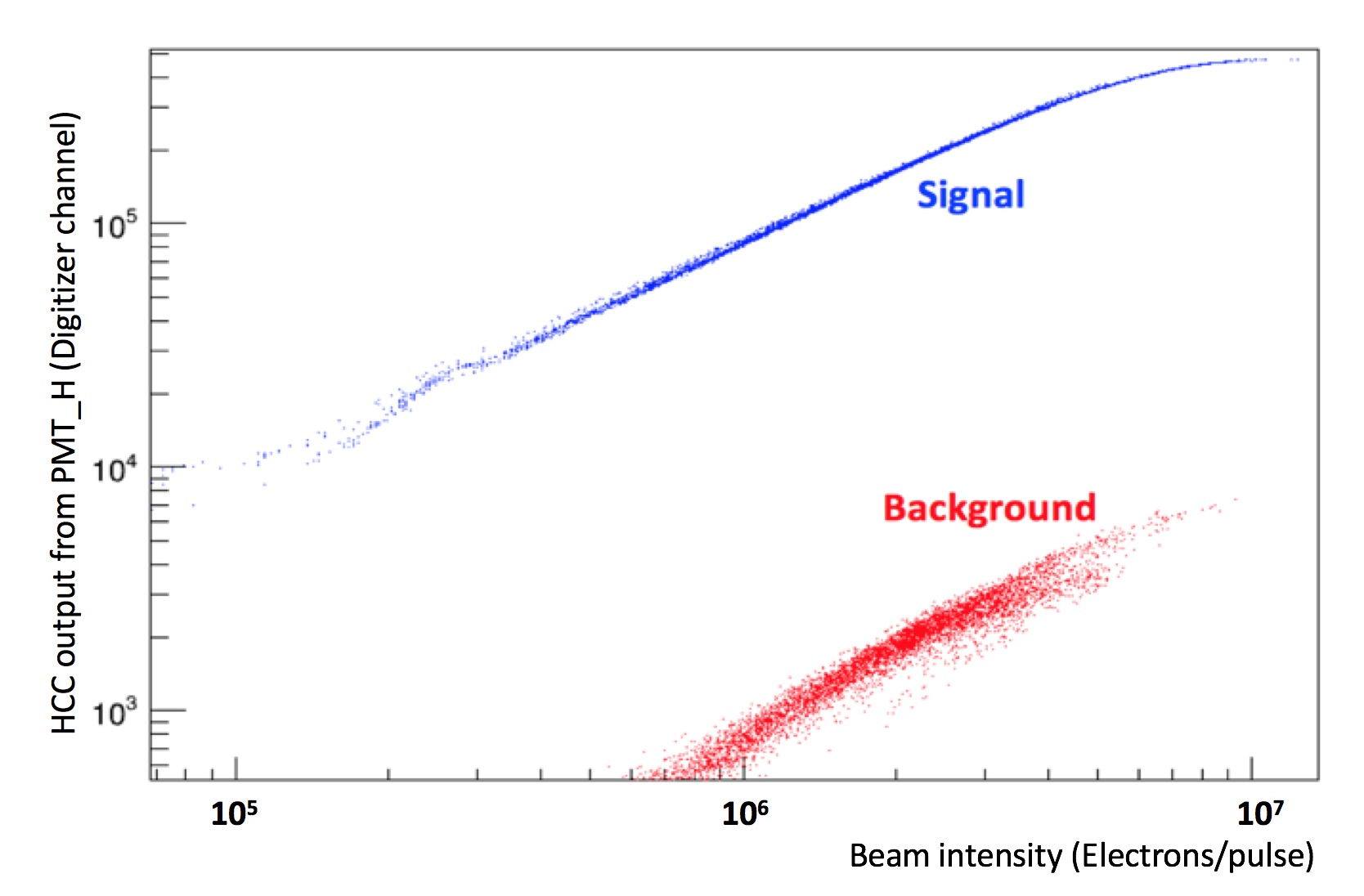}
\figcaption{\label{f15} The background data is obtained by moving HCC 3 cm away from the beam line. In this case, only secondary particles can hit on HCC. Both beam signal and background have a linear relationship with the beam intensity, the ratio of signal to background is about 1.7\%.}
\end{center}

\item[d)]Calibration Error:
Because the FC has a large size, perfect shielding is nearly impossible. Primary electrons may hit on FC directly and lead to an overestimation of beam intensity. The large fluctuation of FC response can also have an influence on the accuracy of IC and PS calibration.
\end{itemize}

\section{Conclusion}
\label{Con}
In this paper, the linearity of LYSO crystal for HCC is studied by measuring the relationship between HCC output and electron beam intensity given by faraday cup, ionization chamber and plastic scintillator. The linearity is better than 10\% from 205 GeV to 1.1$\times$10$^{3}$ TeV (or from 17 GeV/cm$^{3}$ to 93 TeV/cm$^{3}$) which meets the requirement of PeV hadron detection. The linear range can be extrapolated to $\pi^{+}$ (MIP, 30MeV) and protons. No obvious saturation effect of HCC is observed in this range. The uncertainty of linearity measurement is mainly dominated by the measurement system and beam monitoring system. 
\\From this work, we get a preliminary conclusion that the LYSO has a good linearity and can be used in HERD CALO for detecting high energy particles. A beam test at CERN SPS, which can supply electrons and protons with energies up to 300 GeV, is planned to verify main performances of HERD CALO.
\section*{Acknowledgement}
\acknowledgments{We are thankful to Jianchao Sun, Xuan Zhang, Xiaofeng Zhang and Zhe Ning for their support.}
\end{multicols}
\vspace{-1mm}
\centerline{\rule{80mm}{0.1pt}}
\vspace{2mm}

\begin{multicols}{2}

\end{multicols}

\end{CJK*}
\end{document}